\documentclass[twocolumn,aps,prl,showpacs]{revtex4}
\usepackage{graphicx}
\def\be{\begin{equation}}
\def\ee{\end{equation}}
\def\bea{\begin{eqnarray}}
\def\eea{\end{eqnarray}}

\begin{document}

\title{Correlations and periodicities in Himalayan tree ring widths and temperature anomalies through
wavelets}

\author{Prasanta K. Panigrahi} \email{prasanta@prl.ernet.in}
\affiliation{ Physical Research Laboratory, Navrangpura, Ahmedabad
380 009, India}
\author{P. Manimaran}
\author{P. Anantha Lakshmi}
\affiliation{School of Physics, University of Hyderabad,
 Hyderabad 500 046, India}

\author{Ram R. Yadav}
\affiliation{Birbal Sahani Institute of Paleobotany, Lucknow, 160
014, India}

\begin{abstract}

We have studied periodicities and  correlation properties of tree
ring width chronology of deodar tree from Joshimath (1584 - 1999
years) and Uttarkashi (1500 - 2002 years) in the western Himalayas
and the pre-monsoon (March-April-May) temperature anomalies (1876
- 2003) relative to 1961 -1990 mean, through wavelet analysis.
Periodic behavior is observed in the tree ring chronology with
periodicity in the form 11, 22, and 42 years. The analysis of the
self-similar nature reveals long-range correlation with a Hurst
exponent, $H >0.5$. These are anti-correlated with the temperature
anomalies. An interesting inversion behavior is observed around
the year $1750$. The power spectral analysis of the time series
corroborate the results of wavelet method.

\end{abstract}
\maketitle

\section{Introduction}

Our understanding of the variability of climate is largely
hampered by limited length of instrumental weather records,
spanning in most cases to past 100 years. High-resolution proxy
climate records, with precise dating control, provide very good
tool to supplement the weather records back by several centuries
and millennia. Of these records, tree rings provide valuable proxy
as annual growth rings can be precisely dated to calendar year of
their formation and the overlapping template of tree ring
chronologies can be calibrated with weather data to hindcast the
climate variables. Such long-term records can be used to
understand the mode of climate variability in a longer
perspective.

The climate dynamics is affected by a large number of factors,
which in turn is reflected  on a variety of proxy records, such as
tree ring widths, ocean and lake deposits etc. The tree ring data
has a much higher resolution as compared to the later ones.
Recently, these two class of data have been combined through the
multi-resolution capability of the wavelets \cite{mob}, for
reconstructing millennial-scale climate variability, in the
northern hemisphere. Multi-centennial variations in temperature,
possibly arising out of natural phenomena, have been inferred from
the above study, which correlates well with a general circulation
model. The variations in temperature naturally introduces
variations in precipitation patterns, which are truthfully
recorded in the tree ring data. At present, global temperatures
are increasing; the last century in particular has seen
substantial variations in temperature and precipitation rates,
which may be arising due to anthropogenic forcing or natural
causes.

The goal of the present article is to employ wavelet transform for
studying ring chronologies' periodicities in tree ring and detect
the presence of self-similar behavior and correlation properties
\cite{mandel,feder}. The fact that, a number of phenomena in
nature reveal these type of behavior and wavelets provide an ideal
tool to find the same, motivates this study. Apart from the
implications of the periodicities, the nature of the self-similar
fractal behavior, in terms of persistence or anti-persistence will
carry long-term physical implication.

In the following section, we outline briefly the basic properties
of the wavelet transform which are useful for the present
analysis. In section III, a description is provided about the tree
ring materials, chronology preparation and climate data. Sec. IV
deals with results and discussions. We finally conclude in sec. V
with a summary and future directions of work.

\section{Wavelet Transform}
Since the early eighties, wavelet transform has emerged as a
powerful tool to analyze transient and time-varying phenomena
\cite{daub,mall}. It has innumerable applications in various
fields, ranging from signal processing, natural sciences,
economics and finance data analysis \cite{chu, ding,
arn1,mani1,mani2,mani3,mani4,pkp1,pkp2}. Wavelet transform
decomposes an input signal into components that depend on position
and scale. We can characterize the input signal by changing the
scale for a particular location.

The wavelet basis functions are localized both in time and
frequency (or position and scale). The wavelets are parameterized
by the scale parameter (dilation parameter) $s>0$, and a
translation parameter $-\infty < a < \infty$, thus the wavelet
basis can be constructed from one single function $\psi(x)$
according to

\be \psi_{s,a}(x) =\psi ( \frac{x-a}{s} ). \ee

Here, $\psi(x)$ is the mother wavelet. Given a function $f(x)$,
the (continuous) wavelet transform is defined as

\be W[f](s,a) = \frac{1}{\sqrt s} \int_{-\infty}^{+\infty}
\psi_{s,a}^* f(x) dx. \ee

Here $\psi^*(x)$ denotes the complex conjugate of $\psi(x)$. In
order for a function $\psi(x)$ to be usable as an analyzing
wavelet, one must demand that it has zero mean. We make use of
continuous Morlet wavelet for studying the periodicity in our
data. The behavior of the global power spectrum is inferred from
$ln |W_{s,a}[f]|^2$ versus scale $s$.

We employ the Daubechies wavelets for finding the self similar
behavior of the time series. In order to be useful for removing
polynomial trends, wavelets belonging to the Daubechies family are
made to satisfy,

\be \int_{-\infty}^{+\infty} x^m \psi(x) dx = 0, ~~ 0 \leq m \leq
n. \ee

Here the upper limit n is related to what is called the order of
the wavelet. We use Daubechies-12 wavelet to find the Hurst
exponent (H), making use of the average wavelet coefficient method
\cite{awc}. This wavelet has been found to be reliable through
calculating Hurst exponent for the time series like Gaussian white
noise and binomial multifractal model, for which the Hurst
exponent is known \cite{mani5}.

In the average coefficient method for estimating the correlation
behavior, one computes,

\begin{equation}
W[f](s)=\langle |W[f](s,a)| \rangle \simeq s^{1/2 + H}.
\end{equation}
W[f](s) is the averaged wavelet energy of all locations for
different scales. Thus for the given time series, the scaling
exponent (1/2 + H) is measured from the log-log plot of W[f](s)
versus scale $s$ through linear fit. The Hurst exponent (H) varies
between $0 < H < 1$. If $ H < 1/2$, the time series possesses
anti-correlation behavior and for $H > 1/2$ long-range correlation
behavior is present. For uncorrelated time series, $H = 1/2$. We
have also analyzed the time series through Fourier power spectral
analysis, where $P(f) \sim f^{-\alpha}$, and $\alpha = 2 H +1$.

\section{ Tree ring materials and chronology preparation}
Tree ring samples in the form of increment cores were collected
from Himalayan cedar trees growing at moisture stressed sites in
Juma near Joshimath and Gangotri in Uttarkashi. Increment borers
were used to extract 4mm diameter cores from trees at 1.4m stem
height from ground. The increment cores were processed to cross
date the sequence of growth rings in trees to exact calendar year
of their formation. The ring widths of precisely dated growth
rings were measured using linear encoder with the accuracy of
0.01mm. Long-term growth trends inherent in trees due to
increasing age and stem girth were removed by standardizing the
ring width measurement series. Individual tree ring width
measurement series were fitted with negative exponential or linear
regression line with negative slope or no slope and indices
calculated as quotient of actual measurement and curve value. The
individual tree series after standardization were averaged using
bi-weight robust estimation of the mean to develop mean chronology
using program ARSTAN. The chronology dynamics assumed to be
climate driven could be used to examine the possible low frequency
modes and how these might have varied over time
\cite{jay1,ram,jay2}. The tree ring series prepared form Gangotri,
Uttarkashi and Jurna, Joshimath in Uttaranchal are shown in Figs.
1 and 2 respectively.

{\it Climate data}: Tree ring chronologies showed strong negative
relationship with pre-monsoon temperature. To calibrate the tree
ring series, we prepared mean pre-monsoon temperature series by
merging temperature anomalies of nine stations (relative to 1961 -
1990 mean) in western Himalaya. The mean temperature series (Fig.
7) is biased by larger station data records beginning from the mid
of $20^{th}$ century except in case of four stations, where it
extends back to beginning of the $20^{th}$ century and even
earlier.

\section{Results and Discussion}
As mentioned earlier, we make use of the Morlet wavelet for
studying the periodicity in the data. The average wavelet
coefficient method is used for finding the correlation behavior
through calculation of Hurst exponent. It is worth mentioning
that, here we are dealing with non-stationary data; wavelets are
well suited for the analysis of this class of data. The wavelet
coefficients are investigated and wavelet amplitude spectrum is
obtained.

Figs. 1 and 2 depict the global power spectra of the tree data
from Uttarkashi and Joshimath, clearly revealing multiple periodic
variations at 11, 22, and 42 years respectively \cite{nara}. Fig.
3, depicts the global power spectrum of temperature records, which
shows an anti-correlation behavior with the tree ring variations
of Figs. 1 and 2. Scalogram of the wavelet coefficients is given
in Fig. 4; where one can observe the periodicities as mentioned
earlier. An interesting inversion is clearly seen around 1570
years.

\begin{figure}
\centering
\includegraphics[width=2.7in]{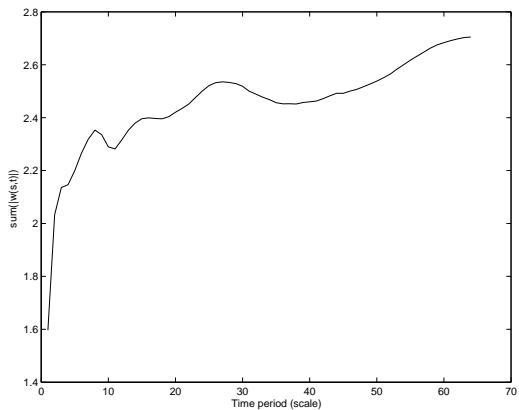}
\caption{The semi-log plot of wavelet power summed over all time
at different scales (time period). One clearly observes the
variations at $11$, $22$ and $42$ years.}
\end{figure}

\begin{figure}
\centering
\includegraphics[width=2.7in]{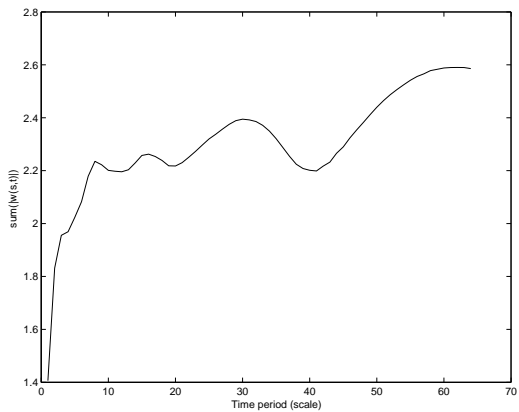}
\caption{the semi-log plot of wavelet power summed over all time
at different scales (time period). Here also one sees the
variations at $11$, $22$ and $42$ years.}
\end{figure}

\begin{figure}
\centering
\includegraphics[width=2.7in]{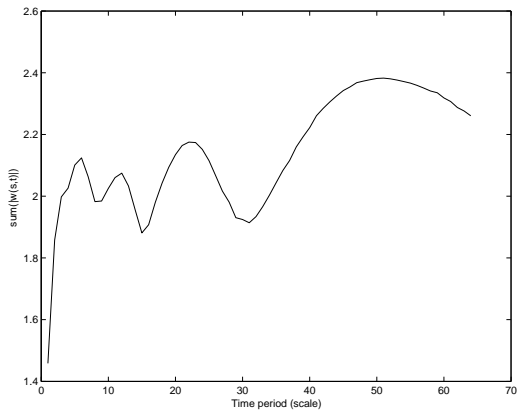}
\caption{The semi-log plot of the wavelet power versus scale (time
period) for the temperature data. An anti-correlation behavior
with the tree ring variations of Figs. $1$ and $2$ is clearly
seen.}
\end{figure}

\begin{figure}
\centering
\includegraphics[width=2.7in]{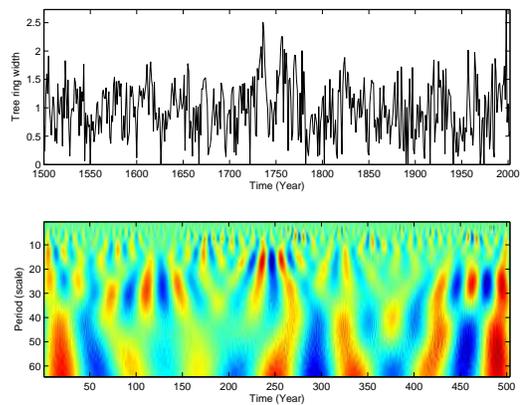}
\caption{(Upper panel) time series of ring width chronology of
deodar from Uttarkashi and (lower panel) scalogram of the above
time series, one can clearly see the inversion around 1570 year
and the periodicities as mentioned above.}
\end{figure}

In Figs. 5, 6 and 7, upper panel (a) shows the time series of
accumulated tree ring and temperature data after subtracting the
mean, whereas (b) shows the power law behavior of the Fourier
spectral analysis of the time series.

\begin{figure}
\centering
\includegraphics[width=3in]{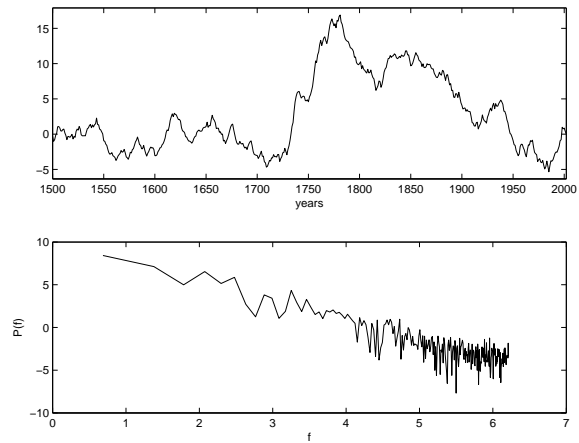}
\caption{Power spectral analysis for ring width chronology of
deodar from Uttarkashi, yielding the scaling exponent $\alpha =
2.45066$.}
\end{figure}

\begin{figure}
\centering
\includegraphics[width=3in]{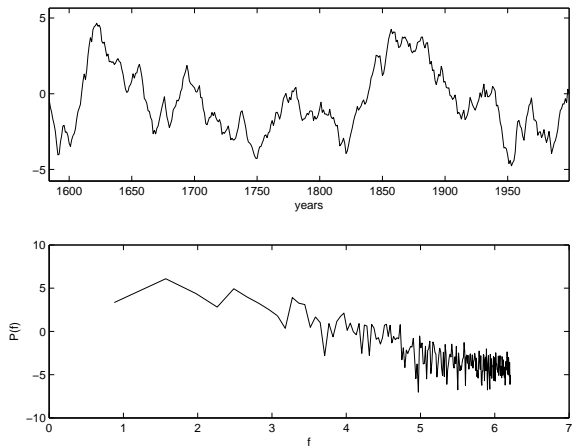}
\caption{Power spectral analysis for ring width chronology of
deodar from Joshimath, giving the scaling exponent $\alpha =
2.4416$.}
\end{figure}

\begin{figure}
\centering
\includegraphics[width=3in]{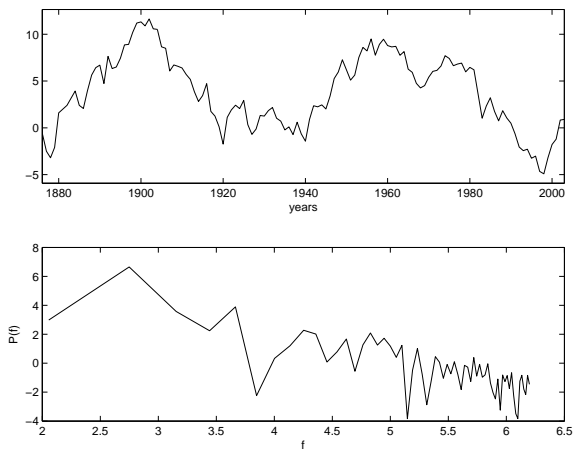}
\caption{Power spectral analysis for pre-monsoon temperature,
giving the scaling exponent $\alpha = 2.34$.}
\end{figure}

We now investigate the correlation properties of tree ring time
series, for which we have used average wavelet co-efficient
method. Hurst exponent, which is a measure of correlation
properties in a time series is computed through Daubechies-12
wavelet. We have found that the tree ring time series possess long
range correlation, the Hurst exponent $H \sim 0.73$ (tree ring
series from UttarKashi), $H \sim 0.74$ (tree ring series from
Joshimath) and for temperature anomalies $H \sim 0.7$. The results
obtained from average wavelet coefficient method is comparable
with the Fourier power spectral analysis $P(s) ~ s^{- \alpha}$
results by the relation $\alpha = 2 H + 1$, keeping in mind the
finite data length. The obtained scaling exponent through Fourier
analysis is $\alpha = 2.45066, H \sim 0.73$ (tree ring series from
Uttarkashi) and $\alpha = 2.4416, H \sim 0.72$ (tree ring series
from Joshimath). Temperature record yields a scaling exponent
$\alpha =2.34$ and the corresponding Hurst exponent $H \sim 0.67$.
This also shows long range correlation behavior.

\section{Conclusion}

From the above obtained results one can clearly see the climate
variations in both time and frequency scales. One observes both
periodic and self-similar processes. Keeping in mind the
non-stationary nature of the time series, the efficacy of the
wavelets in extracting the above behavior is clearly seen. This is
due to the localization and multi-resolution ability of the
wavelets. As expected, there is an anticorrelation between the
tree-widths and temperature anomalies. Surprisingly, the
self-similar behavior yields long-range correlation. This aspect
needs to be further studied carefully in conjunction with other
related data sets, since long-range correlation carries
significant physical implications. In particular, the nature of
these correlations as a function of time is of deep interest. Some
of these studies are currently under progress and will be reported
elsewhere.

\end{document}